\documentclass[reprint,
superscriptaddress,
 amsmath,amssymb,
 aps,
prl
]{revtex4-2}

\usepackage{graphicx}
\usepackage{dcolumn}
\usepackage{bm}
\usepackage{chemformula}
\usepackage[version=4]{mhchem}
\usepackage{xcolor}
\usepackage{braket}

\begin{document}


\title{Bond Anharmonicity in a Caged Molecule Probed by Vibronic Core Spectroscopy}

\author{Connor J.\ B.\ Fields}
\email{connor.fields@nottingham.ac.uk}
\affiliation{School of Physics and Astronomy, University of Nottingham, Nottingham NG7 2RD, United Kingdom}

\author{Bang~Huynh}
\affiliation{New College, University of Oxford, Oxford OX1 3BN, United Kingdom}

\author{Filipe~Junqueira}
\author{Tyler~James}
\affiliation{School of Physics and Astronomy, University of Nottingham, Nottingham NG7 2RD, United Kingdom}

\author{Aleksandra~Foerster}
\affiliation{National Institutes of Health, 9000 Rockville Pike, Bethesda, Maryland 20892, U.S.A.}

\author{Ilya~Popov}
\affiliation{School of Chemistry, University of Nottingham, Nottingham NG7 2RD, United Kingdom}

\author{James Kerfoot}
\affiliation{Nanoscale and Microscale Research Centre, University of Nottingham, Nottingham NG7 2RD, United Kingdom}
\author{Graham Rance}
\affiliation{Nanoscale and Microscale Research Centre, University of Nottingham, Nottingham NG7 2RD, United Kingdom}

\author{Victor~Por\'ee}
\author{Alessandro~Nicolaou}
\affiliation{Synchrotron SOLEIL, L'Orme des Merisiers, Saint-Aubin, BP 48, 91192 Gif-sur-Yvette Cedex, France}

\author{David~Duncan}
\author{Tien-Lin~Lee}
\affiliation{I09 Beamline, Diamond Light Source, Harwell Science and Innovation Campus, Didcot OX11 0DE, United Kingdom}

\author{Gabriela~Hoffman}
\author{Mark~C.~Walkey}
\author{Elizabeth~S.~Marsden}
\author{Malcolm~H.~Levitt}
\author{Richard~J.~Whitby}
\affiliation{School of Chemistry \& Chemical Engineering, University of Southampton, Southampton SO17 1BJ, United Kingdom}

\author{Michael~W.~George}
\author{Elena~Besley}
\affiliation{School of Chemistry, University of Nottingham, Nottingham NG7 2RD, United Kingdom}
\author{Brian~Kiraly}
\author{James~O'Shea}
\author{Philip~Moriarty}
\email{philip.moriarty@nottingham.ac.uk}
\affiliation{School of Physics and Astronomy, University of Nottingham, Nottingham NG7 2RD, United Kingdom}

\date{\today}

\begin{abstract}
X-ray absorption spectroscopy and resonant inelastic X-ray scattering at the nitrogen K-edge show that the N$\equiv$N bond in \ch{N2@C60} is substantially more anharmonic than in free \ch{N2}, with the ground-state ($X^1\Sigma_g^+$) anharmonic constant $\hbar\omega_e x_e$ a factor of two larger ($3.8\pm0.3$~meV versus $\sim 1.8$~meV) despite an essentially unchanged harmonic frequency. The same enhancement is found across all three vibronic potentials accessible to the experiment---the electronic ground state, the core-excited intermediate state, and the spectator-decay final state---pointing to a common origin rather than an effect specific to one electronic configuration. High-level coupled-cluster calculations of the \ch{N2@C60} potential energy surface, however, find no significant change in anharmonicity upon encapsulation, ruling out a simple static distortion of the N$\equiv$N bonding potential by the fullerene cage. We propose instead that the dramatic increase in $\hbar\omega_e x_e$ above the gas-phase value -- which, taken literally, would imply an unphysical halving of the N$_2$ dissociation energy - reflects the breakdown of the single-molecule Morse description, as previously invoked to explain anomalous anharmonicities in 
N$_2$ clusters, and supported by the observation of shifts in the fullerene host's Raman modes.
\end{abstract}

\maketitle

 

Endohedral fullerenes enclose an atomic or molecular guest within a hollow carbon cage, providing a unique form of nanoscale confinement that isolates the encapsulated species from its surroundings while leaving it chemically intact~\cite{EndoFull_BOOK_Yang2014,Levitt2013,Bacanu2025}. For species that interact only weakly with the buckminsterfullerene cage, the ground-state electronic structure is comparable to that in the gas phase -- there is minimal orbital hybridisation with, or charge transfer to, the surrounding buckyball~\cite{Bacanu2025,Bloodworth2022,Fields2025,Krachmalnicoff2016,Levitt2013}. The intracage degrees of freedom are therefore well isolated, providing a novel platform to study the coupling of a localised quantum state to the external environment~\cite{Zurek2003,Schlosshauer2019}. A prominent example is \ch{N@C60}, in which the encapsulated nitrogen atom retains its atomic quartet ground state ($^4S_{3/2}$)~\cite{Mauser1997,Greer2000} and exhibits an exceptionally long electron-spin coherence time~\cite{AlmeidaMurphy1996,Harneit2007}, motivating proposals for endofullerene-based qubit architectures~\cite{Harneit2002,Harneit2017,Pinto2020}. 

The extent to which the intracage isolation is dynamically complete, however, remains an open question. The anharmonicity of the guest's intramolecular potential is, in principle, a sensitive probe of the host-guest coupling dynamics: it captures the variation in the bond potential away from its minimum, and is therefore responsive to interactions that can leave the static electronic structure essentially unperturbed. Confinement effects -- in particular, the quantisation (and coupling) of rotational and translational motion -- have been studied extensively by inelastic neutron scattering~\cite{Horsewill2010,Horsewill2012}, terahertz~\cite{Jafari2022,Jafari2025} and infrared spectroscopy~\cite{Ge2011,Melentev2021}, nuclear magnetic resonance (NMR)~\cite{Chen2013,Mamone2014,Levitt2013,Bacanu2025}, and via computational and analytical calculations~\cite{Bacic2018, Felker2020,Rashed2019}. Moreover, it is now well established that the guest-cage interaction potential can have significant anharmonic character~\cite{Jafari2022,Jafari2023,Jafari2025}. By contrast, the anharmonicity of the intramolecular potential of the encapsulate itself has received comparatively little attention, and its sensitivity to the host-guest coupling dynamics has not, to the best of our knowledge, previously been explored experimentally.

Smucker and P\'erez-R\'{\i}os~\cite{Smucker2025} recently calculated the spectroscopic constants of \ch{N2@C60} from a rotationally-averaged potential energy surface, finding only a modest reduction in the first anharmonic correction term, $\hbar\omega_e x_e$, relative to the free molecule. Our own frozen-cage coupled-cluster calculations (Supp. Mat.) likewise find no significant change in anharmonicity upon encapsulation. High-level quantum chemical theory therefore predicts minimal change, at most, in the anharmonic constant. As we show here, however, this is at odds with the results of vibronic core spectroscopy of \ch{N2@C60}. We suggest that the discrepancy between theory and experiment arises from coupling between the N$\equiv$N stretch and the modes of the C$_{60}$ host: coupled oscillators do not retain the vibrational levels of their isolated components~\cite{Redlich1941, Heller1980, Doronin2011}.For \ch{N2@C60}, this implies that the anharmonicity observed in core spectroscopy is a property of the coupled guest--host system, not solely of the N$\equiv$N bond. 

Near-edge X-ray absorption fine structure (NEXAFS) spectroscopy and resonant inelastic X-ray scattering (RIXS) provide element-specific access to the vibronic structure of the guest across multiple electronic states: by tuning to the N K-edge, the response of the encapsulated N$_2$ is isolated from that of the C$_{60}$ host. This combination of element specificity and access to excited-state potentials is not available from infrared, Raman (see End Matter) or neutron-based techniques. \ch{N2@C60} is particularly 
well suited to this approach: the nitrogen molecule is closed-shell 
with no orbital hybridisation with the cage~\cite{EndoFull_N2aC60CompStudy_Slanina2006, Smucker2025}, and 
the N$\equiv$N triple bond produces well-resolved vibronic features 
that have allowed the intramolecular potential to be determined with 
high precision in both gas-phase and adsorbed 
N$_2$~\cite{N2NEX_firstNEX_Chen1989,N2NEX_N2ResPES_Neeb1994,N2NEX_BetterResNtheory_Feifel2004,N2_gasRIXS_Kjellsson2021, Doronin2011}.

N$_2$@C$_{60}$ (Fig. 1(a)) was synthesised using molecular surgery~\cite{HoffmanThesis2021} and deposited as thick films/multilayers onto Si(111) or Ag(111) substrates via thermal sublimation under ultrahigh vacuum conditions (Supp. Mat.). RIXS measurements were acquired using the AERHA spectrometer of the \mbox{SEXTANTS} beamline at the \mbox{SOLEIL} synchrotron~\cite{Beamlines_AERHA_Chiuzbaian2014,N2NEX_SEXTANTSBeam_Sacchi2013}. The incident photon energy was scanned through the $\mathrm{N}~1\mathrm{s}\rightarrow\pi^*$ resonance (Fig. 1(b)), and the emitted-photon spectrum (spanning a 12 eV range below the elastic line) was recorded at each excitation energy. (We adopt a localised core-hole description throughout and denote the core-excited state as $1\mathrm{s}^{-1}\pi^*$.)  Complementary NEXAFS spectroscopy (upper inset to Fig. 1(c)) and core-level and valence-band photoemission (Supp. Mat.) were carried out at Beamline I09 at the Diamond Light Source~\cite{I09info_Lee2018}. All NEXAFS and RIXS spectra acquired at SOLEIL and the Diamond Light Source were calibrated by aligning the fitted 0-0 component of the N~$1\mathrm{s}\rightarrow \pi^*$ vibronic progression to the literature gas-phase value of 400.88~eV~\cite{N2NEX_BetterResNtheory_Feifel2004, N2_gasRIXS_Kjellsson2021}. Further details of sample preparation and photoemission characterisation are provided in the Supplemental Material.

\begin{figure}[t!]
    \centering
    \includegraphics[width=0.48\textwidth]{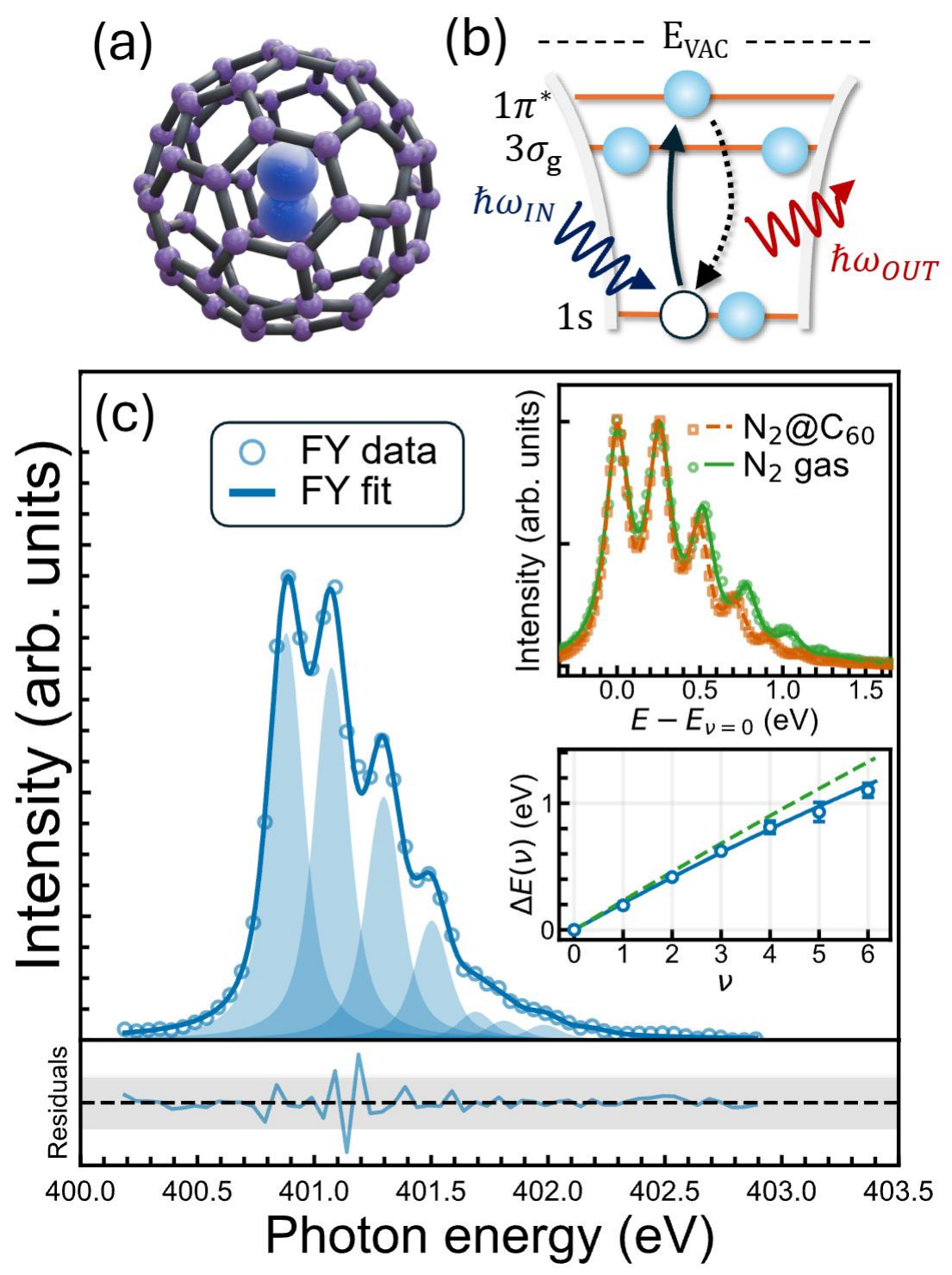}
\caption{\textbf{N K-edge NEXAFS of \ch{N2@C60}.}
\textbf{(a)} Structure of \ch{N2@C60} (C, purple; N, blue).
\textbf{(b)} Energy-level diagram for the N~$1\mathrm{s}\rightarrow\pi^*$ resonance (black solid arrow) and a decay to the electronic ground state (dashed arrow) with the emission of an X-ray photon of energy $\hbar\omega_{\mathrm{OUT}}$. The molecule may be left vibrationally excited, so that, in this case, the energy loss $\hbar\omega_{\mathrm{IN}} - \hbar\omega_{\mathrm{OUT}}$ equals the vibrational excitation energy of the $X^1\Sigma_g^+$ state. Other decay channels that leave the molecule electronically excited are discussed below.   
\textbf{(c)} Fluorescence-yield (FY) NEXAFS spectrum fitted with Voigt components assigned to vibrational levels of the $1\mathrm{s}^{-1}\pi^*$ core-excited state. Lower panel: residuals; grey band: $\pm2\sigma$. Upper inset: total-electron-yield spectrum of a thick \ch{N2@C60} film and the spectrum of gas-phase N$_2$, both measured at I09 (Diamond Light Source) and aligned at $v=0$. Lower inset: fitted peak energies, with Morse fit (solid line) and the progression expected for the core-excited state of gas-phase N$_2$ (green dashed line).}
    \label{fig:NEXAFS}
\end{figure}

The upper inset of Fig. 1(c) is a comparison of nitrogen K-edge NEXAFS spectra measured for a multilayer of N$_2$@C$_{60}$ (see Supp.~Mat. for photoemission spectra of this sample) and for gaseous N$_2$, both acquired on the same beamline (I09, Diamond Light Source) where the additional anharmonicity for N$_2$@C$_{60}$ is clear from the lower energies of the higher-lying vibronic peaks. The nitrogen K-edge NEXAFS spectrum shown in the main panel of Fig.~\ref{fig:NEXAFS}(c) was obtained from a RIXS map (at SEXTANTS) by integrating the emitted-photon intensity over the measured emission-energy window at each incident photon energy~\cite{NKEdge_TFY_Jaklevic1977,N2NEX_PFY_Boyko2014}. A clear vibronic progression associated with the $1\mathrm{s}^{-1}\pi^*$ core-excited state is again observed. We have fitted the spectrum using a series of Voigt components assigned to successive vibrational levels of the $1\mathrm{s}^{-1}\pi^*$ state. In the lower inset to Fig 1(c) open circles show the fitted vibronic peak positions plotted as a function of $v$, the vibrational quantum number, while the solid line shows the Morse energy eigenvalue 
expression~\cite{Morse1929, PESVib_MorseAnalytical_Dahl1988}
\begin{equation}
E(v) =
E_{00}
+ \hbar\omega_{e}\left(v+\frac{1}{2}\right)
- \hbar\omega_{e}x_{e}\left(v+\frac{1}{2}\right)^2 ,
  \label{eq:morse_levels}
  \end{equation}
fitted to those points and plotted continuously as a guide to the eye. Here, $E_{00}$ is the minimum of the potential, $\hbar\omega_e$ is the harmonic vibrational term, and $\hbar\omega_e x_e$ is the anharmonic correction. Fitting the NEXAFS peak positions as a quadratic function of $v+\tfrac{1}{2}$ yields a harmonic energy of $\hbar\omega_e = 230\pm10$~meV and an anharmonic constant, $\hbar\omega_e x_e$, of $6\pm2$~meV. The harmonic term is close to the gas-phase value for the core-excited state, while the anharmonicity is substantially larger than the gas-phase value of approximately $2.1$~meV~\cite{N2_gasRIXS_Kjellsson2021}. For comparison, the inset also shows the vibrational level spacings calculated from the known gas-phase harmonic and anharmonic constants~\cite{N2NEX_BetterResNtheory_Feifel2004, N2_gasRIXS_Kjellsson2021} (dashed green line).

\begin{figure}
\centering
\includegraphics[width=0.49\textwidth]{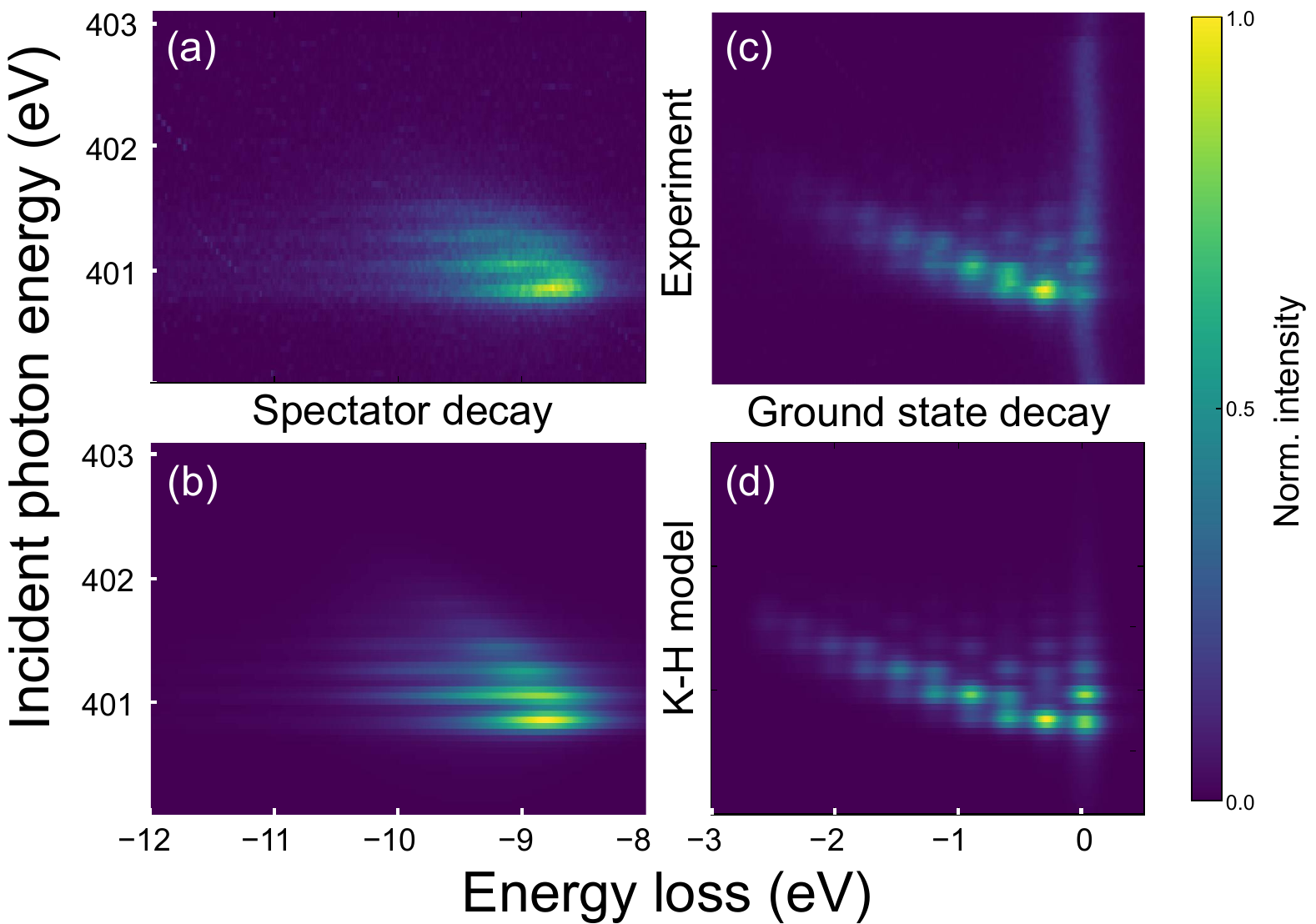}
\caption{\textbf{Nitrogen K-edge RIXS maps and Kramers--Heisenberg model of N$_2$@C$_{60}$.}
Experimental RIXS intensity maps are shown for (a) the spectator-decay channel and (c) the ground-state energy-loss region, with incident photon energy scanned through the N~$1\mathrm{s}\rightarrow 1\pi^*$ resonance. The corresponding Kramers--Heisenberg (K-H) simulations are shown in (b) and (d), calculated using Morse potentials for the ground, core-excited, and final electronic states.}
\label{fig:rixsmap}
\end{figure}

Figure~\ref{fig:rixsmap} shows the complete nitrogen K-edge RIXS response of N$_2$@C$_{60}$ as the incident photon energy is scanned through the N~$1\mathrm{s}\rightarrow \pi^*$ resonance. At higher energy loss (Fig.~\ref{fig:rixsmap}(a)), a set of features is observed due to the spectator-decay channel. In this process, the core hole that is present for NEXAFS measurements is filled while the electron promoted into the $\pi^*$ orbital remains in the valence shell. This leaves the molecule in a valence-excited $3\sigma_g^{-1}\pi^*$ final state. In contrast to the dispersive spectator features, the electronically elastic decay channel (Fig.~\ref{fig:rixsmap}(c)) returns the molecule to the electronic ground state $X^1\Sigma_g^+$ (dashed arrow in Fig. 1(b)), producing a vibrational progression at approximately fixed energy loss. Figs.~\ref{fig:rixsmap}(b,d) are the corresponding simulated RIXS maps for the spectator- and ground-state-decay features, calculated using the Kramers-Heisenberg method described in the Appendix. 

Integrating the RIXS intensity over the incident-energy axis yields 
one-dimensional energy-loss projections (Fig.~3). As the spectator-decay projection (Fig.~3(a)) is dispersive, individual vibronic components comprising the 
spectator envelope are shifted in energy loss with incident photon energy. The 
projection is therefore constructed from the K-H model as a sum of 
NEXAFS-weighted vibronic components evaluated at their dispersion-shifted 
positions. By contrast, the ground-state projection (Fig.~3(b)) involves 
an electronically elastic channel that is non-dispersive so that each peak falls at  an approximately fixed energy loss; the vibrational level spacings of $X^1\Sigma_g^+$ are therefore visible directly in the integrated spectrum. The vibrational constants extracted from the K-H analysis are listed in Table~\ref{tab:KHparams}. Note that while the harmonic vibrational frequency is essentially unaffected, there is a pronounced increase in the anharmonic term. For the electronic ground state, the fitted value of $3.8\pm0.3$~meV is more than twice the gas-phase value of approximately $1.8$~meV~\cite{HuberHerzberg1979, N2_gasRIXS_Kjellsson2021, Wang2023}. The same trend is observed for the excited states.

\begin{figure*}[t!]
    \centering
    \includegraphics[width=0.85\textwidth]{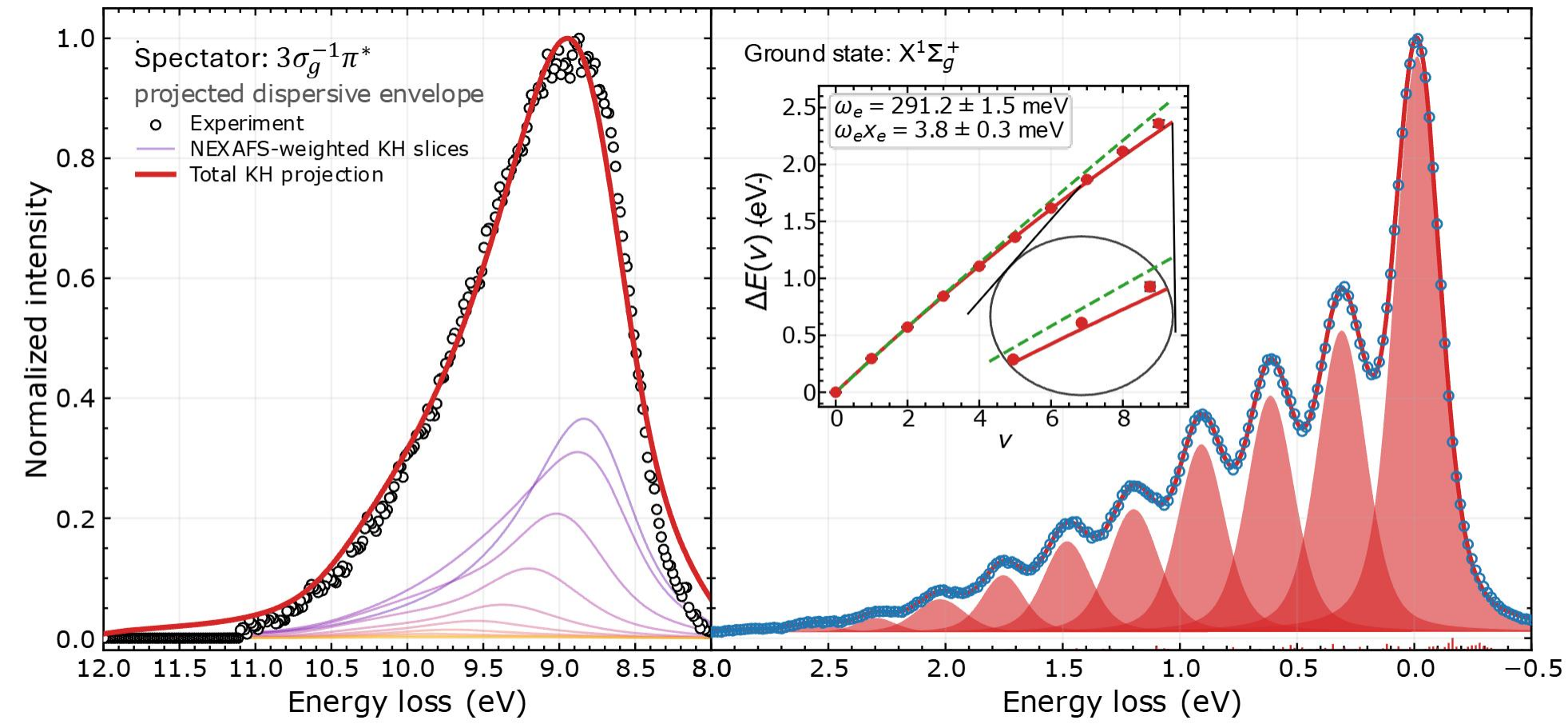}
    \caption{\textbf{Energy-loss projections of the spectator-decay and ground-state RIXS channels.}
    \textbf{(a)} Projection of the spectator-decay region, obtained by 
    summing the RIXS intensity over the incident-energy range of the 
    N~$1\mathrm{s}\rightarrow 1\pi^*$ resonance. The experimental spectrum is 
    compared with the total Kramers--Heisenberg projection and the 
    individual NEXAFS-weighted slices used to reproduce the dispersive 
    spectator-decay envelope. 
    \textbf{(b)} Projection of the ground-state decay region, showing 
    the vibrational progression of the $X^1\Sigma_g^+$ state. The fitted 
    inelastic peaks correspond to increasing vibrational excitation of 
    the N$\equiv$N stretch. Inset: fitted ground-state vibrational 
    excitation energies plotted against $v$, 
    with the solid line showing the Morse-level fit used to extract 
    $\hbar\omega_e$ and $\hbar\omega_e x_e$. Green dashed line: gas-phase N$_2$.} 
    \label{fig:projections}
\end{figure*}

To investigate the microscopic origin of the enhanced anharmonicity, 
the N$\equiv$N potential energy surface was computed for both free and 
C$_{60}$-encapsulated N$_2$ using the domain-based local pair natural orbital 
coupled-cluster method, DLPNO-CCSD(T)/def2-SVP~\cite{Riplinger2013, Riplinger2013b, Weigend2005} (Supp.~Mat.).  Although the calculations 
reproduced the harmonic frequency of free N$_2$ to within a few 
percent of the experimental value, neither the equilibrium nor the displaced configurations yield any significant increase in anharmonicity upon encapsulation. This 
insensitivity of the computed anharmonicity to the guest--host 
geometry, combined with the near-unchanged $\hbar\omega_e$ seen in 
both experiment and calculation, suggests that the enhanced 
anharmonicity is not driven by a static modification of the N$\equiv$N 
potential by the cage environment.

Since $D_e = (\hbar\omega_e)^2 / 
(4\hbar\omega_e x_e)$~\cite{Morse1929, Doronin2011, Denisov2025}, a doubling of $\hbar\omega_e x_e$ with an 
essentially unchanged $\hbar\omega_e$ implies a halving of the N$_2$
dissociation energy. This is clearly physically 
untenable given the lack of hybridisation of N$_2$ and 
C$_{60}$ orbitals~\cite{Smucker2025, EndoFull_N2aC60CompStudy_Slanina2006}. As such, the factor of two increase in the anharmonicity constant for encaged N$_2$ as compared to that of its gas-phase counterpart cannot reflect a genuine static modification of the N$\equiv$N bond potential by the cage environment alone, and must instead arise from the breakdown of the single-molecule Morse description when applied to a coupled guest--host system. Anomalously large changes in $\hbar\omega_e x_e$ have similarly been 
reported previously for nitrogen clusters: Doronin~\textit{et~al.}~\cite{Doronin2011} observed that free N$_2$ clusters exhibit a ground-state anharmonicity approximately half that of the gas-phase single-molecule value, with $\hbar\omega_e$ essentially unchanged. Drawing on the work of Redlich~\cite{Redlich1941} and Heller and Gelbart~\cite{Heller1980}, Doronin \textit{et al.} attributed the ``anomalous'' anharmonicity constants they measured to interactions between 
neighbouring N$\equiv$N bonds within a coupled-oscillator framework. 
Our observation of dramatically altered (albeit increased rather than decreased)
$\hbar\omega_e x_e$ constants with near-unchanged $\hbar\omega_e$ is reminiscent of that of Doronin \textit{et al.}~\cite{Doronin2011} and similarly points
to a breakdown of the Morse description when the 
N$\equiv$N stretch is coupled to additional degrees of freedom. 

\begin{table}[b!]
\begin{ruledtabular}
\begin{tabular}{lcc}
State
    & $\hbar\omega_e$ (meV)
    & $\hbar\omega_e x_e$ (meV) \\
$X^1\Sigma_g^+$ (ground)
    & $291.0 \pm 1.5$
    & $3.8 \pm 0.3$ \\
$1\mathrm{s}^{-1}\pi^*$ (core-excited)
    & $230 \pm 10$
    & $6 \pm 2$\\
$3\sigma_g^{-1}\pi^*$ (spectator)
    & $240 \pm 10$
    & $6 \pm 4$ \\
\end{tabular}
\end{ruledtabular}
\caption{\textbf{Harmonic and anharmonic constants for  ground, core-excited, and spectator states.} Vibrational constants were obtained from the projection-constrained fitting procedure described in the End Matter section. The ground-state values were extracted from the near-elastic energy-loss projection, the core-excited values from the NEXAFS spectrum (Fig. 1), and the spectator-state values from the spectator-decay projection and full K-H map.}
\label{tab:KHparams}
\end{table}

Direct experimental evidence of coupling between the encaged N$_2$ molecule and the C$_{60}$ host is provided by Raman 
spectroscopy (Fig.~4 and Table II, End Matter). Several low-frequency cage modes are blue-shifted relative to their positions for the empty C$_{60}$ molecule; higher frequency modes are red-shifted. These shifts --- in opposite 
directions for modes of radial and tangential character, consistent 
with observations for Ar@C$_{60}$~\cite{Cimpoesu2011} --- provide strong evidence for cage-guest coupling. By the symmetry of any coupling Hamiltonian, the cage modes must also perturb the N$\equiv$N vibrational states (in a 
manner that is invisible to the frozen-cage calculations).

Importantly, the coupling implied by the Raman shifts need not be 
large to produce the observed apparent anharmonicity enhancement. 
The anharmonicity constant $\hbar\omega_e x_e$ is determined by 
the second finite difference of the vibrational level energies. A 
perturbation $\delta E(v)$ to the level energies modifies the 
fitted anharmonicity by
\begin{equation}
\delta(\hbar\omega_e x_e) = -\frac{1}{2}
\frac{\delta E(v+2) - 2\,\delta E(v+1) + \delta E(v)}{\Delta v},
\end{equation}
which depends only on the curvature of $\delta E(v)$ 
with respect to $v$. For a perturbation of the form 
\mbox{$\delta E(v) = -\epsilon v^2$} --- the simplest form with non-zero 
curvature, and physically natural for any coupling to an external 
degree of freedom that grows with vibrational amplitude --- the 
second finite difference is $-2\epsilon$, giving 
$\delta(\hbar\omega_e x_e) = +\epsilon$. To double 
$\hbar\omega_e x_e$ therefore requires $\epsilon = 
\hbar\omega_e x_e = 1.78$~meV. At $v = 3$, this corresponds to 
a perturbation of $\epsilon v^2 = 16$~meV, which is less than 
2\% of $E(v=3) \approx 855$~meV --- entirely consistent with a weak van der Waals-mediated guest--host interaction ~\cite{Smucker2025, EndoFull_N2aC60CompStudy_Slanina2006}. A fully coupled quantum treatment of the guest--host vibrational 
eigenstates --- analogous to that developed by 
Felker and Ba\v{c}i\'{c}~\cite{Felker2020} but extended to resolve the anharmonic structure of the vibrational ladder --- is needed to account in full for our
observations.\\

\begin{acknowledgments}
\noindent \textbf{Acknowledgements.} We gratefully acknowledge both the SOLEIL synchrotron and Diamond Light Source (DLS) for the provision of beamtime via awards 20230252 and 99240058 (SOLEIL) and allocations SI-31574-1 and SI31574-2 (DLS). C.F. also thanks DLS for the award of PhD studentship STU0425. Endofullerene synthesis was supported by EPSRC grants EP/P009980/1 and  EP/T004320/1. P.M. thanks the EPSRC for an Established Career Fellowship (EP/T033568/1). Python code for the handling, fitting, and plotting of RIXS, NEXAFS, and photoemission data was developed with the assistance of Open AI's ChatGPT (ver 4.0 and 4.5) and Anthropic's Claude Opus 5.5. The authors specified the analysis, checked the code and its outputs against the raw data, and verified all reported values.
\end{acknowledgments}

\noindent\textit{Data availability}---The data that support the findings of this article are openly available at \url{https://repository.nottingham.ac.uk/handle/123456789/96741} (DOI:~\url{https://doi.org/10.17639/nott.47998}.)

\bibliography{N2C60}
\onecolumngrid
\newpage
\section*{End Matter}
\twocolumngrid
\subsection{Appendix A: Kramers--Heisenberg modelling.} To extract quantitative information from the RIXS maps, the spectra were modelled using a Kramers--Heisenberg description of resonant scattering~\cite{N2RIXS_Eloss_Kotani2001,DeGroot2008,N2_gasRIXS_Kjellsson2021}. In this formalism, RIXS is treated as a second-order process in which the system is first excited from the electronic ground state into a core-excited intermediate state and then decays into an allowed final state. Within the Franck--Condon approximation, the electronic and nuclear parts of the transition matrix element can be separated~\cite{Stohr1991,NEXAFS_VibronicBOOK_Kato2021}. Each electronic state is therefore described by its own potential-energy curve, supporting a corresponding ladder of vibrational levels and nuclear wavefunctions. Through vibronic coupling, the change in electronic configuration during excitation and decay allows the system to be projected onto different vibrational levels and electronic states. These vibrational-state populations modify the emitted-photon energies and intensities, producing the  features observed in the RIXS maps. Since the model considers a single dominant electronic excitation and decay channel, the electronic contribution can be absorbed into an overall intensity factor, leaving the relative RIXS intensities to be determined by Franck--Condon transitions~\cite{PESVib_MorseAnalytical_Dahl1988,N2_gasRIXS_Kjellsson2021}.

For each final vibrational state $|f\rangle$, the state-resolved scattering intensity was calculated from the coherent sum over intermediate vibrational levels~\cite{N2RIXS_Eloss_Kotani2001,N2_gasRIXS_Kjellsson2021},
\begin{equation}
I_{\mathrm{scat}}(E_{\mathrm{ex}}) \propto
\left|
\sum_{n}
\frac{
\langle f|n\rangle
\langle n|g\rangle
}{
E_{\mathrm{ex}}-\left(E_n-E_g\right)+i\Gamma_c/2
}
\right|^2 ,
\label{eq:KH_scattering}
\end{equation}
where $E_{\mathrm{ex}}$ is the incident photon energy, $|g\rangle$ is the initial vibrational state on the electronic ground-state potential, $|n\rangle$ denotes vibrational levels of the core-excited intermediate-state potential, and $|f\rangle$ denotes vibrational levels of the final-state potential. The terms $\langle n|g\rangle$ and $\langle f|n\rangle$ are Franck--Condon overlaps between the corresponding nuclear wavefunctions, while $\Gamma_c$ accounts for the finite lifetime of the core-excited state. 

The full RIXS intensity map was then obtained by placing each state-resolved contribution at the corresponding energy loss,
\begin{equation}
I_{\mathrm{RIXS}}(E_{\mathrm{ex}},E_{\mathrm{loss}}) \propto
\sum_f
I_{\mathrm{scat}}(E_{\mathrm{ex}})
\delta\left(
E_{\mathrm{loss}}-\left(E_f-E_g\right)
\right).
\label{eq:KH_RIXS}
\end{equation}
The Dirac delta function enforces energy conservation by placing each final vibrational state at its corresponding energy loss. In the numerical model used to generate the maps in Fig.~\ref{fig:rixsmap}, this delta function was replaced by a peak broadening function, while the finite incident-energy resolution was included by convolution along the excitation-energy axis. In this form, Eq.~\ref{eq:KH_scattering} retains the coherent interference between pathways through different intermediate vibrational levels, while Eq.~\ref{eq:KH_RIXS} maps the resulting final-state intensities onto the energy-loss axis of the calculated RIXS spectrum.

The values used to construct the Kramers--Heisenberg maps in Fig.~\ref{fig:rixsmap} are listed in Table~\ref{tab:KHparams}. These parameters were obtained through a multistep fitting procedure rather than by a direct unconstrained fit to the full two-dimensional maps. In brief, the RIXS-derived NEXAFS spectrum and the energy-loss projections of the ground-state and spectator-decay regions were first used to determine the vibronic peak positions, widths, and relative energy offsets. The resulting parameters were then used to generate the full Kramers--Heisenberg simulations shown in Fig.~\ref{fig:rixsmap}. 

The discrete vibronic transitions generated by the Kramers--Heisenberg model were broadened to account for the experimental resolution before comparison with the data. In the incident-energy direction, the linewidth reflects both the finite lifetime of the core-excited state and the excitation-energy resolution, and was therefore described using a Voigt profile. In the energy-loss direction, the linewidth is dominated by the spectrometer response. This response was also modelled empirically using a Voigt function. This does not imply an additional lifetime-broadening contribution along the energy-loss axis; rather, the Voigt profile provides a flexible description of the non-trivial spectrometer response, which is not captured adequately by either a purely Gaussian or purely Lorentzian lineshape~\cite{N2NEX_VoigtWidths_Olivero1977,N2RIXS_RIXSWidths_Ketenoglu2015,N2RIXS_RIXSWidths2_im2018}. The same broadening convention was used for the RIXS-derived NEXAFS spectrum and the full RIXS maps.

\begin{figure*}
    \centering
    \includegraphics[width=0.8\textwidth]{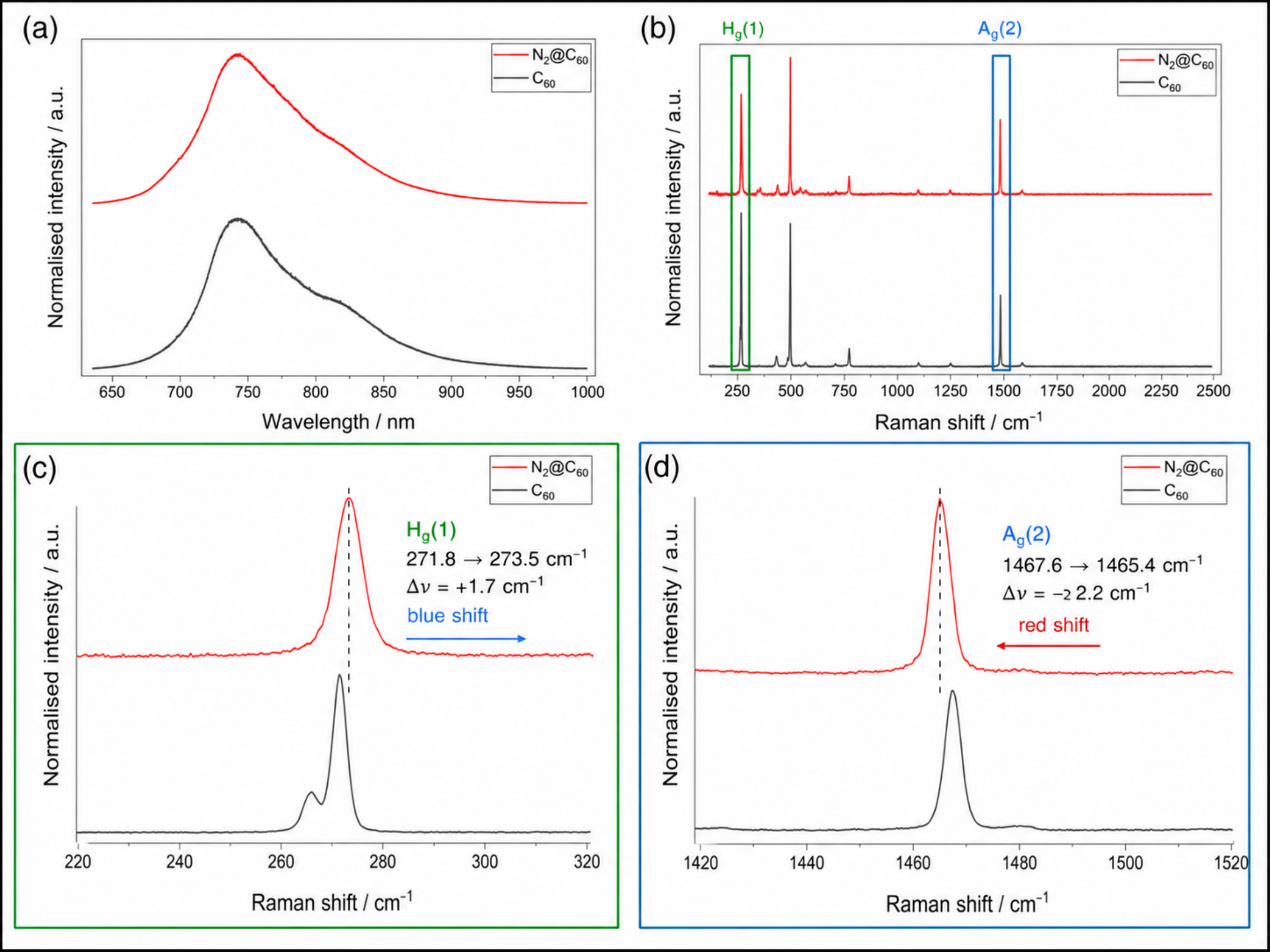}
    \caption{\textbf{Micro-photoluminescence and Raman spectroscopy of C$_{60}$ and N$_2$@C$_{60}$.}
    \textbf{(a)} Normalised micro-photoluminescence spectra of bulk C$_{60}$ and N$_2$@C$_{60}$, shown in grey and red, respectively.
    \textbf{(b)} Corresponding micro-Raman spectra over the full measured Raman-shift range. The highlighted regions indicate the $H_{\mathrm{g}}(1)$ and $A_{\mathrm{g}}(2)$ cage modes enlarged in panels (c) and (d), respectively.
    \textbf{(c)} Enlargement of the low-frequency $H_{\mathrm{g}}(1)$ mode, which shifts from 271.8~cm$^{-1}$ in C$_{60}$ to 273.5~cm$^{-1}$ in N$_2$@C$_{60}$, corresponding to a blue shift of $\Delta\tilde{\nu}=+1.7$~cm$^{-1}$.
    \textbf{(d)} Enlargement of the high-frequency $A_{\mathrm{g}}(2)$ mode, which shifts from 1467.6~cm$^{-1}$ in C$_{60}$ to 1465.4~cm$^{-1}$ in N$_2$@C$_{60}$, corresponding to a red shift of $\Delta\tilde{\nu}=-2.2$~cm$^{-1}$.
    The spectra are normalised and vertically offset for clarity. The mode-dependent shifts demonstrate that encapsulated N$_2$ measurably perturbs the vibrational potential of the fullerene cage, providing evidence for guest--host coupling.}
    \label{fig:Raman}
\end{figure*}

\subsection{Appendix B: Mode coupling evidenced through Raman spectroscopy.}

Micro-photoluminescence ($\mu$PL) and micro-Raman spectroscopy
($\mu$RS) measurements were performed on bulk C$_{60}$ and
N$_2$@C$_{60}$ using a commercial optical spectrophotometer
(LabRAM Evolution HR, HORIBA, France). The instrument was equipped
with a 633~nm He--Ne laser (Pacific Lasertec, USA) and a 785~nm
semiconductor diode laser (Innovative Photonic Solutions, USA),
operated at incident powers of 0.07~mW and 0.6~mW, respectively. The scattered light was collected using a
$100\times$, 0.9 numerical-aperture metallurgical objective
(Olympus, Japan) and dispersed using a 600~lines~mm$^{-1}$
diffraction grating. Edge filters with a low-frequency cut-off of
approximately 50~cm$^{-1}$ were employed, and the confocal pinhole
was set to 200~$\mu$m. The single-pixel spectral dispersions at
zero Raman shift were better than 0.8 and 0.5~cm$^{-1}$ for the
633~nm and 785~nm configurations, respectively. The spectrometer was
calibrated before measurement using a Standard Pro-Raman
Calibration Objective. Samples were dispersed in methanol and
deposited by drop casting.

\begin{table}
\centering
\caption{Assigned Raman-active cage modes of empty C$_{60}$ and
N$_2$@C$_{60}$.}
\label{tab:raman_modes}
\begin{tabular}{lccc}
\hline
Mode &
C$_{60}$ / cm$^{-1}$ &
N$_2$@C$_{60}$ / cm$^{-1}$ &
$\Delta\tilde{\nu}$ / cm$^{-1}$ \\
\hline
$H_{\mathrm{g}}(1)$ & 271.8  & 273.5  & +1.7 \\
$H_{\mathrm{g}}(2)$ & 431.8  & 437.5  & +5.7 \\
$A_{\mathrm{g}}(1)$ & 495.7  & 495.9  & +0.2 \\
$G_{\mathrm{g}}(2)$ & 567.2  & 569.4  & +2.2 \\
$H_{\mathrm{g}}(4)$ & 772.1  & 771.7  & -0.4 \\
$H_{\mathrm{g}}(5)$ & 1099.6 & 1098.6 & -1.0 \\
$H_{\mathrm{g}}(6)$ & 1250.1 & 1246.0 & -4.1 \\
$H_{\mathrm{g}}(7)$ & 1422.7 & 1419.4 & -3.3 \\
$A_{\mathrm{g}}(2)$ & 1467.6 & 1465.4 & -2.2 \\
$H_{\mathrm{g}}(8)$ & 1573.7 & 1572.4 & -1.3 \\
\hline
\end{tabular}
\end{table}

Direct experimental evidence of the interaction between the encapsulated N$_2$ molecule and the C$_{60}$ host is provided by the Raman spectroscopy data presented in Fig.~\ref{fig:Raman} and Table~\ref{tab:raman_modes}. Several low-frequency cage modes are blue-shifted following encapsulation, including the
$H_{\mathrm{g}}(1)$ and $H_{\mathrm{g}}(2)$ modes, by 1.7 and 5.7~cm$^{-1}$, respectively. In contrast, several
higher-frequency modes are red-shifted, including $H_{\mathrm{g}}(6)$, $H_{\mathrm{g}}(7)$ and $A_{\mathrm{g}}(2)$, by 4.1, 3.3 and 2.2~cm$^{-1}$, respectively. These mode-dependent changes demonstrate that the
presence of N$_2$ measurably modifies the cage vibrational potential and provide evidence for guest--host coupling.

\end{document}


\title{Supplemental Material for:\\
Bond Anharmonicity in a Caged Molecule \\Probed by Vibronic Core Spectroscopy}

\author{Connor J.\ B.\ Fields}
\email{connor.fields@nottingham.ac.uk}
\affiliation{School of Physics and Astronomy, University of Nottingham, Nottingham NG7 2RD, United Kingdom}

\author{Bang~Huynh}
\affiliation{New College, University of Oxford, Oxford OX1 3BN, United Kingdom}

\author{Filipe~Junqueira}
\author{Tyler~James}
\affiliation{School of Physics and Astronomy, University of Nottingham, Nottingham NG7 2RD, United Kingdom}

\author{Aleksandra~Foerster}
\affiliation{National Institutes of Health, 9000 Rockville Pike, Bethesda, Maryland 20892, U.S.A.}

\author{Ilya~Popov}
\affiliation{School of Chemistry, University of Nottingham, Nottingham NG7 2RD, United Kingdom}

\author{James Kerfoot}
\affiliation{Nanoscale and Microscale Research Centre, University of Nottingham, Nottingham NG7 2RD, United Kingdom}
\author{Graham Rance}
\affiliation{Nanoscale and Microscale Research Centre, University of Nottingham, Nottingham NG7 2RD, United Kingdom}

\author{Victor~Por\'ee}
\author{Alessandro~Nicolaou}
\affiliation{Synchrotron SOLEIL, L'Orme des Merisiers, Saint-Aubin, BP 48, 91192 Gif-sur-Yvette Cedex, France}

\author{David~A.~Duncan}
\author{Tien-Lin~Lee}
\affiliation{I09 Beamline, Diamond Light Source, Harwell Science and Innovation Campus, Didcot OX11 0DE, United Kingdom}

\author{Gabriela~Hoffman}
\author{Mark~C.~Walkey}
\author{Elizabeth~S.~Marsden}
\author{Malcolm~H.~Levitt}
\author{Richard~J.~Whitby}
\affiliation{School of Chemistry and Chemical Engineering, University of Southampton, Southampton SO17 1BJ, United Kingdom}

\author{Michael~W.~George}
\author{Elena~Besley}
\affiliation{School of Chemistry, University of Nottingham, Nottingham NG7 2RD, United Kingdom}
\author{Brian~Kiraly}
\author{James~O'Shea}
\author{Philip~Moriarty}
\email{philip.moriarty@nottingham.ac.uk}
\affiliation{School of Physics and Astronomy, University of Nottingham, Nottingham NG7 2RD, United Kingdom}

\maketitle

\section{Sample preparation} \noindent Milligram quantities of N$_2$@C$_{60}$ powder, synthesised via molecular surgery\cite{Bloodworth2022} with subsequent purification by HPLC and sublimation, were provided by the Southampton group (Whitby \textit{et al.})  For the NEXAFS measurements at the Diamond Light Source (beamline I09) shown in Fig. 1(c) of the main paper, the endofullerene sample was prepared as follows. Following degassing for a minimum period of twelve hours, a multilayer film of N$_2$@C$_{60}$ was grown by sublimation at a nominal evaporator temperature of 670$\pm$20~K onto a clean Ag(111) surface held at 180 K. (Although not discussed here, submonolayer and monolayer coverages of N$_2$@C$_{60}$ that produced a ($2\sqrt{3} \times 2\sqrt{3}$)R$30^{\circ}$ low energy electron diffraction (LEED) pattern were also deposited and spectroscopically analysed prior to the multilayer studies.) 

We note that although the Ag(111) surface prior to the deposition of N$_2$@C$_{60}$ was free of an O 1s photoemission signal, in the initial stages of the first deposition of N$_2$@C$_{60}$ a small oxygen contribution was observed, which we attribute to the presence of interstitial water and/or oxidised fullerene at the surface of the material in the evaporator crucible.  Growth of a sufficiently thick N$_2$@C$_{60}$ film ``buried'' the oxygen contaminant such that it was located outside the escape depth of the photoelectrons. Alternatively, evaporating off the multilayer (by heating the Ag(111) substrate) followed by sputter-anneal cycles restored a clean, fullerene-free Ag(111) surface. Subsequent depositions of N$_2$@C$_{60}$ from the same evaporator showed no evidence of the contaminant. To circumvent beam damage, all measurements at I09 were made at a sample temperature of 180 K and with the undulator flux deturned by approximately a factor of ten from the peak output.

A different sample preparation strategy was used for the RIXS measurements at SOLEIL (SEXTANTS beamline). In that case, samples were prepared \textit{ex situ} and transported to the beamline end-station through air. To ensure that the films were microns thick at a minimum, all of the $\sim$ 1 mg N$_2$@C$_{60}$ source material was sublimed from the evaporator. The samples were subsequently transferred into the SEXTANTS end-station UHV chamber (base pressure $\sim$ 5$\times 10^{-9}$ mbar) and cooled using a liquid-helium flow cryostat to approximately 40~K during the measurements. For the data shown in Fig. 2 of the main paper four RIXS maps were acquired and averaged to improve signal-to-noise ratio.

\section{Valence band and core-level photoemission spectroscopy}

\noindent Core-level and valence-band photoelectron spectra were recorded at the soft X-ray branch of beamline I09 at the Diamond Light Source~\cite{I09info_Lee2018}, using a Scienta EW4000 hemispherical analyser. Binding energies are referenced to the Fermi level of the Ag(111) substrate (extracted from a fit of a Fermi-Dirac function). Rough estimates of film thickness based on the ratio of C 1s to Ag 3d intensity suggest that the valence band spectrum shown in Fig.~S1 arises from a film that is of order 4 to 5 layers thick. Given the substrate temperature during deposition (180 K), the growth was unlikely to be ideally layer-by-layer, however. In line with this suggestion, the lack of a ``drop to baseline'' between the HOMO and HOMO-1 peaks in Fig.~S1 arises from photoemission from the Ag(111) substrate. Nonetheless, the valence band spectrum is wholly characteristic of the C$_{60}$ cage and, as for many other endofullerenes synthesised by molecular surgery\cite{EndoFull_MoEFXSWEXP_Jarvis2021,Fields2025,Sufyan2025}, shows that encapsulation leaves the frontier orbitals of the cage essentially unchanged.

\subsection{N~1s}
\begin{figure}[b!]
  \centering
  \includegraphics[width=0.98\textwidth]{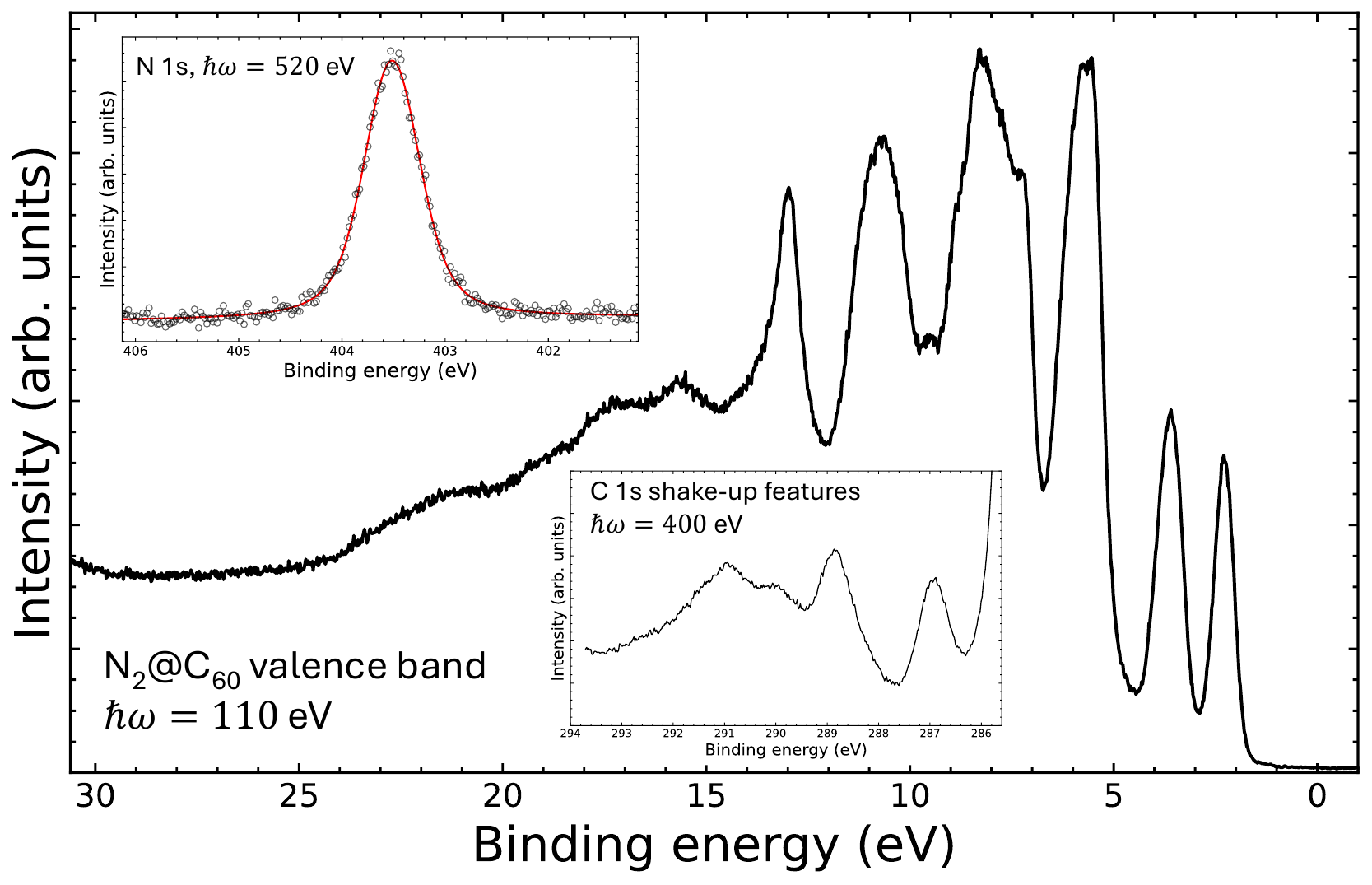}
  \caption{\textbf{Photoelectron spectroscopy of a multilayer film of N$_2$@C$_{60}$ on Ag(111).} The main panel is the valence band spectrum measured with a photon energy of 110 eV. The prominent, characteristic leading features at $\sim$ 2.3 eV and $\sim$ 3.7 eV below the Fermi level (0 eV binding energy) arise from the highest occupied molecular orbital (HOMO) and HOMO-1 of the C$_{60}$ cage. \textbf{Lower~inset:} C 1s shake-up features. The shake-up spectrum is essentially indistinguishable from that of empty C$_{60}$.~\textbf{Upper~inset:} Open circles: N 1s core-level spectrum measured with a photon energy of 520 eV. A single-component Voigt fit is shown as a red line.}
  \label{fig:photoemission}
\end{figure}

\noindent The N~1s spectrum (Fig.~\ref{fig:photoemission}, upper inset; $h\nu = 520$) consists of a single, symmetric line, well described by one Voigt component on a linear background. The fit gives a binding energy of $403.5\pm0.1$~eV and a FWHM of 0.65~eV (Gaussian 0.47~eV, Lorentzian 0.29~eV). The two N atoms of the encapsulated molecule are therefore chemically equivalent, consistent with negligible orbital mixing with the cage. Assuming a typical work function of $\approx4.6$--$4.8$~eV for a C$_{60}$ multilayer on Ag(111), the vacuum-referenced N~1s binding energy is $\approx408.2$--$408.4$~eV. This is $\approx1.5$--$1.7$~eV below that of free N$_2$ ($\approx409.9$~eV), a shift consistent with polarisation screening of the core hole by the surrounding cage and film.

\subsection{C~1s shake-up spectrum}
\noindent The C~1s shake-up spectrum (Fig.~\ref{fig:photoemission}, lower inset) for N$_2$@C$_{60}$ shows a number of well-defined satellite features at energies above the bulk C~1s line. As we have previously discussed in the context of noble gas endofullerenes\cite{Sufyan2025}, these coincide with the shake-up energies of empty C$_{60}$~\cite{Enkvist1993,Maxwell1994}. Because the satellite energies directly reflect excitations between the cage's frontier orbitals, this agreement, together with the valence-band spectrum, again indicates that the electronic structure of the cage is not measurably perturbed by the encapsulated N$_2$.

We also note that the C~1s spectrum was recorded at an earlier, thinner stage of N$_2$@C$_{60}$ deposition than that from which the N~1s spectrum was acquired. Its main line contained, in addition to the bulk component, a component shifted by $\approx0.5$~eV to lower binding energy arising from first-layer molecules chemisorbed on Ag(111) (see, for example, the SI for \cite{Fields2025} and references therein). This interface contribution would at most broaden the lowest-energy satellite and does not affect the comparison with empty C$_{60}$.

\section{Coupled-cluster calculations for N$_2$ and N$_2$@C$_{60}$}

\noindent We analysed \NtwoCsixty{} using high-level coupled-cluster calculations in order to obtain a more accurate description of the potential energy surface than was previously possible with DFT methods.

\subsection{Isolated \ch{N2} molecule}\noindent First, we constructed the PES for isolated \ch{N2} at several levels of theory and analysed the data using both Morse and polynomial fits (Table~\ref{tab:N2}). Although the Morse fits visually reproduced the PES curves well, we found that the extracted vibrational parameters depended strongly on the fitting range. We therefore decided to proceed with a polynomial fit. With the polynomial fits, we could monitor convergence by increasing the polynomial order. In the fit we kept the number of PES points fixed, while varying the degree of the polynomial expansion. Each value of $n$ corresponds to a polynomial of order $n$. The polynomial fits were constructed such that the lower-order coefficients, $c_2$ and $c_3$, remained close to the values obtained from the initial Morse estimate. This constraint was important because unconstrained high-order polynomial fits were very flexible and produced unphysical values of $c_2$ and $c_3$. As a result, the converged values matched experiment reasonably well, which gave us more confidence in the reliability of the fit.

\begin{table}[b!]
  \caption{Vibrational parameters of gas phase \ch{N2} obtained from Morse and polynomial fits at different levels of theory.}
    \centering
  \label{tab:N2}
  \small\setlength{\tabcolsep}{4pt}
  \begin{tabular}{l cc cc cc}
    \toprule
    & \multicolumn{2}{c}{$\omega_e$ (cm$^{-1}$)}
    & \multicolumn{2}{c}{$\wexe$ (cm$^{-1}$)}
    & \multicolumn{2}{c}{$x_e$} \\
    \cmidrule(lr){2-3}\cmidrule(lr){4-5}\cmidrule(lr){6-7}
    Method & Morse & Polynomial & Morse & Polynomial & Morse & Polynomial \\
    \midrule
    DLPNO-CCSD/def2-SVP        & 2463.19 & \pmval{2437}{23} & 13.65 & \pmval{13.80}{0.3} & 0.006 & \pmval{0.006}{0.0001} \\
    DLPNO-CCSD/aug-cc-pVDZ     & 2396.00 & \pmval{2442}{21}  & 16.00 & \pmval{14.12}{0.4} & 0.006 & \pmval{0.006}{0.0001} \\
    DLPNO-CCSD(T)/def2-SVP     & 2403.29 & \pmval{2423}{23} & 14.23 & \pmval{14.13}{0.3} & 0.006 & \pmval{0.006}{0.0001} \\
    DLPNO-CCSD(T)/aug-cc-pVDZ  & 2329.51 & \pmval{2340}{20} & 14.60 & 14.30              & 0.006 & \pmval{0.006}{0.0002} \\
    CCSD(T)/aug-cc-pVDZ        & 2319.66 & \pmval{2323}{15} & 14.75 & \pmval{14.98}{0.3} & 0.006 & \pmval{0.006}{0.0001} \\
    CCSD(T)/aug-cc-pVTZ        & 2340.78 & \pmval{2345}{18} & 14.45 & \pmval{14.65}{0.3} & 0.006 & \pmval{0.006}{0.0001} \\
    CCSD(T)/aug-cc-pVQZ        & 2355.75 & \pmval{2363}{16} & 14.41 & \pmval{14.52}{0.2} & 0.006 & \pmval{0.006}{0.0001} \\
    \midrule
    {Experiment} & \multicolumn{2}{c}{2358.52} & \multicolumn{2}{c}{14.35} & \multicolumn{2}{c}{0.006} \\
    \bottomrule
  \end{tabular}
\end{table}

The CCSD(T)/aug-cc-pVQZ level of theory produced results that were in the closest agreement with the experimental data. However, calculations for \NtwoCsixty{} at this level of theory become computationally demanding. Therefore, we used the DLPNO-CCSD(T)/def2-SVP approach, which still provides a reliable approximation and is considered a high-level quantum-chemical approach.\footnote{DLPNO is a method that retains only the important virtual orbitals associated with each occupied electron pair contributing to the correlation energy, rather than the full virtual space used in canonical CCSD(T).}

\section*{\NtwoCsixty}

\begin{figure}[htbp]
  \centering
  \includegraphics[width=0.7\textwidth]{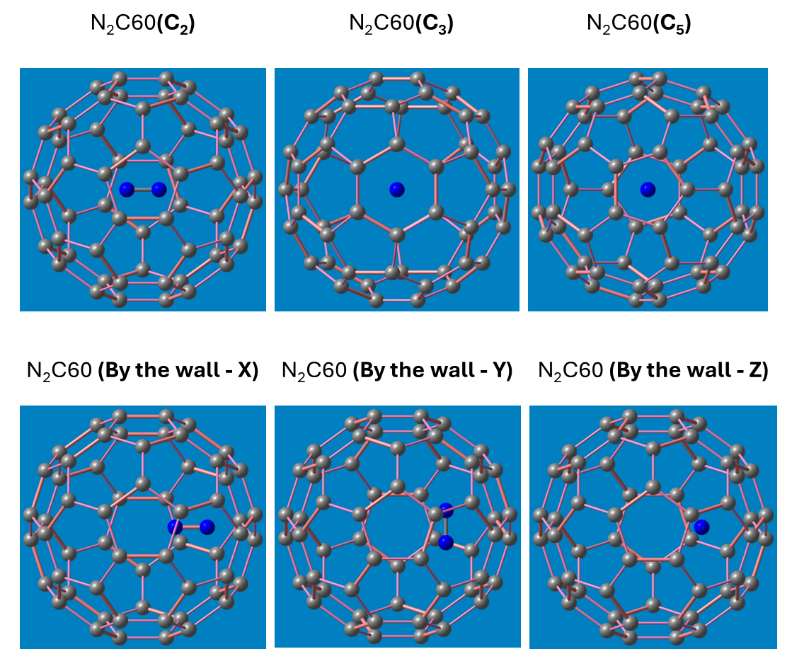}
  \caption{Tested structures of \NtwoCsixty.}
  \label{fig:structures}
\end{figure}

\begin{figure}[htbp]

  \includegraphics[width=\textwidth]{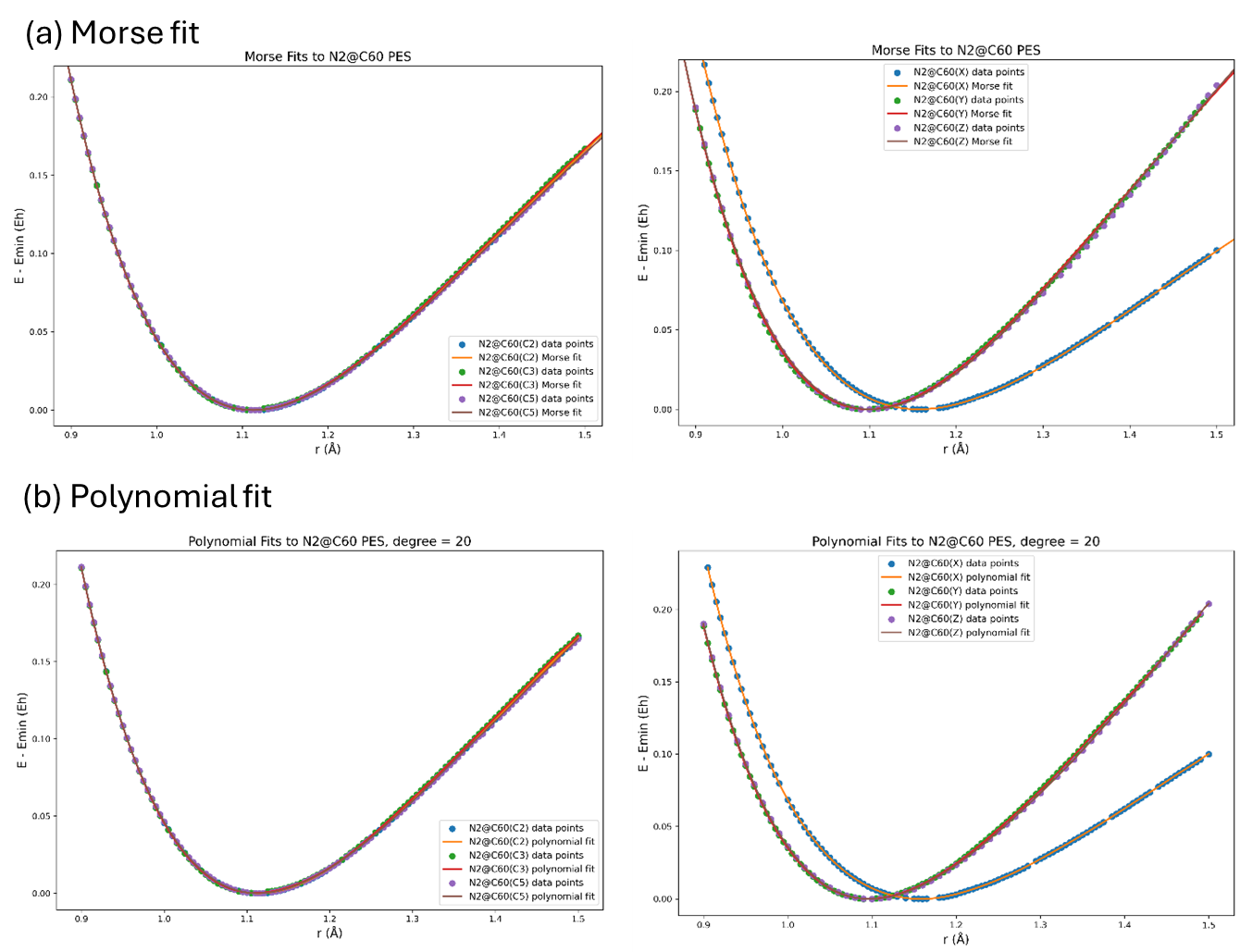}
  \caption{Potential energy surfaces (PES) for \NtwoCsixty{} calculated at the DLPNO-CCSD(T)/def2-SVP level of theory together with fitted (a) Morse potentials and (b) polynomial fits. The left panels show the $C_2$, $C_3$, and $C_5$ symmetry orientations of the fullerene cage, while the right panels correspond to configurations in which the \ch{N2} molecule is positioned close to the fullerene wall along the $X$, $Y$, and $Z$ orientations. Again, both fits appear to be good; however, the vibrational parameters calculated using the polynomial fit (Table~\ref{tab:N2C60}) were more consistent than those obtained from the Morse fit.}
  \label{fig:PES}
\end{figure}

\noindent Using DLPNO-CCSD(T)/def2-SVP we calculated the PES for \ch{N2} encapsulated within \ch{C60} for three orientations, aligned with the cage’s $C_2$, $C_3$ and $C_5$ axes (Fig.~\ref{fig:structures}). In all DLPNO-CCSD calculations, the cage was frozen and only the N-N bond length was varied to trace out the PES. Although this takes into account any coupling in the electronic structure between N$_2$ and the cage (e.g. charge transfer or hybridisation), it cannot account for vibrational coupling. A full vibrational coupling at the DLPNO-CCSD level (involving VSCF/VCI based on mode-coupled PES generated by the Adaptive Density-Guided Approach (ADGA) with DLPNO-CCSD as the underlying electronic-structure method) is currently too computationally expensive to run. Hence, our mitigation was to physically move the encapsulated N$_2$ molecule to different positions and orientations inside the cage to probe for any vibrational coupling effects. Thus, to investigate whether stronger cage--molecule interactions could induce noticeable changes in the harmonic frequency $\omega_e$ or the anharmonicity, we placed the \ch{N2} molecule closer to the fullerene wall in three orientations ($X$, $Y$, and $Z$) (Fig.~\ref{fig:structures}) and also calculated the PES for this configuration (Table~\ref{tab:N2C60}). Figure~\ref{fig:PES} shows the PES calculated at the DLPNO-CCSD(T)/def2-SVP level of theory together with the fitted (a) Morse potentials and (b) polynomial fits for \NtwoCsixty{} in the $C_2$, $C_3$, and $C_5$ orientations (left-hand side), and for \ch{N2} positioned close to the fullerene wall in the $X$, $Y$, and $Z$ orientations (right-hand side). All calculations were performed using the DLPNO-CCSD(T)/def2-SVP method.

\begin{table}[htbp]

  \caption{Vibrational parameters of \NtwoCsixty{} obtained from Morse and polynomial fits. Level of theory: DLPNO-CCSD/def2-SVP}
  \label{tab:N2C60}
  \small\setlength{\tabcolsep}{4.5pt}
  \begin{tabular}{l cc cc cc}
    \toprule
    & \multicolumn{2}{c}{$\omega_e$ (cm$^{-1}$)}
    & \multicolumn{2}{c}{$\wexe$ (cm$^{-1}$)}
    & \multicolumn{2}{c}{$x_e$} \\
    \cmidrule(lr){2-3}\cmidrule(lr){4-5}\cmidrule(lr){6-7}
    Structure & Morse & Polynomial & Morse & Polynomial & Morse & Polynomial \\
    \midrule
    \NtwoCsixty{} ($C_2$)          & 2393.18 & \pmval{2378.31}{17} & 16.72 & \pmval{13.55}{0.4} & 0.007 & \pmval{0.006}{0.0002} \\
    \NtwoCsixty{} ($C_3$)          & 2394.31 & \pmval{2376.89}{17} & 16.75 & \pmval{14.65}{0.3} & 0.007 & \pmval{0.006}{0.0001} \\
    \NtwoCsixty{} ($C_5$)          & 2387.23 & \pmval{2375.98}{20} & 16.80 & \pmval{14.38}{0.7} & 0.007 & \pmval{0.006}{0.0003} \\
    \NtwoCsixty{} (by the wall, $X$) & 1987.12 & \pmval{2029.69}{14} & 13.72 & \pmval{15.95}{1.6} & 0.007 & \pmval{0.008}{0.0007} \\
    \NtwoCsixty{} (by the wall, $Y$) & 2612.20 & \pmval{2567.43}{11} & 15.94 & \pmval{15.32}{0.8} & 0.006 & \pmval{0.006}{0.0003} \\
    \NtwoCsixty{} (by the wall, $Z$) & 2450.56 & \pmval{2481.18}{12} & 17.06 & \pmval{17.18}{1.3} & 0.007 & \pmval{0.007}{0.0005} \\
    \midrule
    {Experiment} & \multicolumn{2}{c}{{2348.70}} & \multicolumn{2}{c}{{30.64}} & \multicolumn{2}{c}{{0.013}} \\
    \bottomrule
  \end{tabular}
\end{table}

Different symmetry orientations of the fullerene cage do not significantly affect either the vibrational frequency or the anharmonicity. If anything, the $X$ orientation in the structures ``\NtwoCsixty{} by the wall'', which is the closest to the fullerene wall, showed the largest change in bond distance and harmonic frequency $\omega_e$, while the anharmonicity changed very little. 

\section{Tip-enhanced Raman spectroscopy measurements}
\label{sec:TERS}

\begin{figure}

    \includegraphics[width=0.95\linewidth]{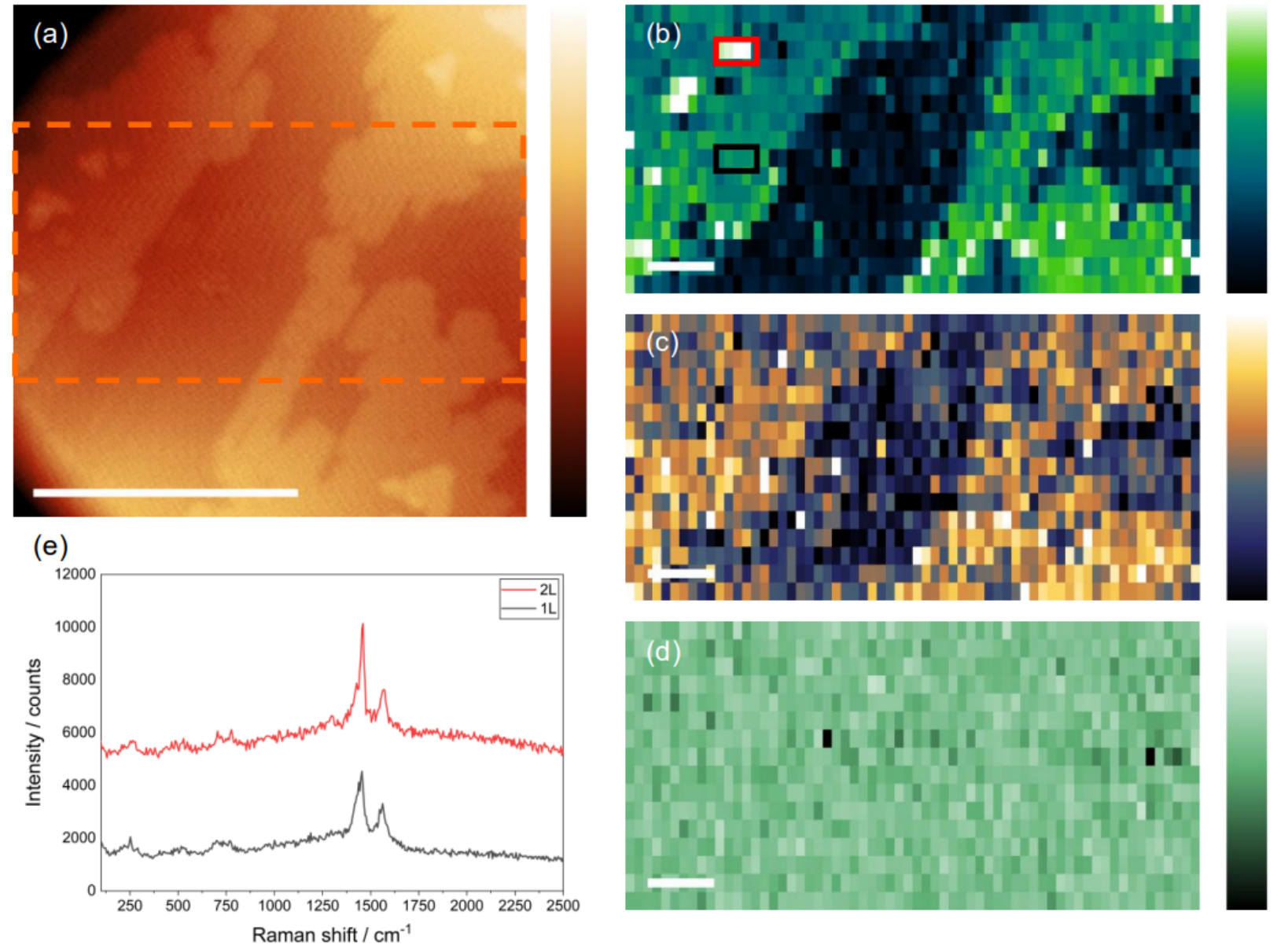}
    \caption{\textbf{AFM and tip-enhanced Raman spectroscopy of a mixed \ch{C60}/\ch{N2@C60} layer on Au.}
    \textbf{(a)} Tapping-mode AFM topographic image showing the formation of monolayer (1L) and bilayer (2L) fullerene domains on the Au surface. The displayed height range is $-5$ to $+5$~nm. The dashed orange box marks the region over which the TERS maps in panels \textbf{(b)}--\textbf{(d)} were acquired.
    \textbf{(b)--(d)} TERS intensity maps obtained by integrating the Raman signal over the spectral ranges
    $1350$--$1500~\mathrm{cm^{-1}}$ \textbf{(b)},
    $1500$--$1650~\mathrm{cm^{-1}}$ \textbf{(c)}, and
    $2200$--$2450~\mathrm{cm^{-1}}$ \textbf{(d)}.
    These ranges contain contributions from the $H_{\mathrm{g}}(7)$ and $A_{\mathrm{g}}(2)$ modes of \ch{C60}, the $H_{\mathrm{g}}(8)$ mode of \ch{C60}, and the $\nu_{\mathrm{N\equiv N}}$ stretching mode of encapsulated \ch{N2}, respectively. The displayed intensity ranges are $0.5\times10^{3}$ to $3.5\times10^{3}$ counts in \textbf{(b)}, $-1.0\times10^{3}$ to $1.0\times10^{3}$ counts in \textbf{(c)}, and $-1.0\times10^{3}$ to $0.5\times10^{3}$ counts in \textbf{(d)}.
    \textbf{(e)} Representative TER spectra acquired at the monolayer and bilayer positions marked by the black and red boxes, respectively, in panel \textbf{(b)}. The spectra are dominated by the $H_{\mathrm{g}}(7)$, $A_{\mathrm{g}}(2)$, and $H_{\mathrm{g}}(8)$ modes of \ch{C60}, whose intensities are generally greater within the bilayer region. No distinct signal attributable to the intramolecular vibration of encapsulated \ch{N2} was resolved. The available spectral resolution was also insufficient to determine small changes in the fullerene-mode peak positions that might otherwise provide indirect confirmation of the presence of \ch{N2} within the \ch{C60} cage.}
    \label{fig:TERS}
\end{figure}

Tip-enhanced Raman spectroscopy (TERS) measurements were performed to investigate whether the spatial distribution of \ch{N2@C60} could be correlated with the topographic and vibrational features of a mixed \ch{C60}/\ch{N2@C60} layer deposited on Au. The resulting AFM and TERS measurements are presented in Fig.~\ref{fig:TERS}. The topographic image in Fig.~\ref{fig:TERS}(a) was acquired using a SmartSPM system (HORIBA, France) operated in tapping mode. During acquisition, the lateral drift was measured to be approximately $0.82~\mathrm{nm\,min^{-1}}$.

Reflection-mode TERS measurements were performed using a commercial LabRAM Nano system (HORIBA, France), comprising the same AFM and optical spectrometer used for the AFM and micro-Raman spectroscopy measurements, respectively. The two instruments were optically coupled using an OmegaScope platform (HORIBA, France) equipped with a $100\times$, 0.7-NA metallurgical objective lens (Mitutoyo, Japan), providing side illumination at an angle of $60^{\circ}$ relative to the sample-surface normal. A 150-lines-$\mathrm{mm^{-1}}$ diffraction grating was used, providing a spectral resolution better than $6.6~\mathrm{cm^{-1}}$ with a pixel-binning value of 2.

Before each measurement, the laser was aligned with the probe apex by scanning the objective position and mapping the intensity of the weak plasmonic signal produced by the tip while it was retracted from the sample surface. TERS measurements were conducted using a 633-nm excitation laser because a greater near-field signal was observed at this wavelength than with the 785-nm laser for the commercial Omni-TERS-SNC-Au probes employed (AppNano, USA). The maximum incident laser power was limited to $0.07~\mathrm{mW}$ to reduce the likelihood of thermal or chemical damage to either the probe or the sample.

Measurements were performed in ``SpecTop'' mode. At each measurement position, the probe was operated in contact mode to collect the combined near-field and far-field Raman signal, $S_{\mathrm{NF+FF}}$. The probe was subsequently operated in tapping mode to collect the far-field-only signal, $S_{\mathrm{FF}}$, and while moving between measurement positions. The Raman contrast factor was defined as
\begin{equation}
C_{\mathrm{R}}
=
\frac{S_{\mathrm{NF+FF}}}{S_{\mathrm{FF}}}-1.
\label{eq:TERS_contrast}
\end{equation}
Using the HORIBA reference sample, comprising carbon nanotubes supported on Au, a typical probe used in this study produced a Raman contrast factor of approximately 15. The spatial resolution, estimated using the same reference sample, was approximately 15~nm.

\bibliography{N2C60_SI}